\documentclass[12pt,fleqn]{article}
\usepackage{epsf}
\begin{document}

\textheight=22cm

\noindent
{\LARGE \bf Fold-Hopf Bursting in a Model for Calcium Signal Transduction}\\

\noindent{{\large Lutz Brusch$^1$, Wolfram Lorenz$^1$, Michal Or-Guil$^1$, 
Markus B\"ar$^1$ and Ursula Kummer$^2$\\}

\noindent
$^1$ Max Planck Institute for the Physics of Complex Systems,\\
N\"othnitzer Str.\ 38, 01187 Dresden, Germany\\
$^2$ Bioinformatics and Computational Biochemistry Group, 
European Media Laboratory, Schloss-Wolfsbrunnenweg 33, 
69118 Heidelberg, Germany\\} 
 
\noindent 6. August 2001


\noindent{\bf Abstract:}\\
We study a recent model for calcium signal transduction. This model displays
spiking, bursting and chaotic oscillations in accordance with experimental 
results. We calculate bifurcation diagrams and study
the bursting behaviour in detail. This behaviour is classified according
to the dynamics of separated slow and fast subsystems. 
It is shown to be of the Fold-Hopf type, a type which was previously
only described in the context of neuronal systems, but not in the
context of signal transduction in the cell. \\  

\section{Introduction}
\label{intro}

Nonlinear dynamics has attracted much attention in several scientific
fields, predominantly in physics. However, oscillations and chaos have 
also been discovered within several levels of abstraction in biology 
and (bio)chemistry. These levels range from 
populations of organisms (predator and prey systems) \cite{beg}, individual
organisms (circadian and ultradian clocks) \cite{edm} down to biochemical reaction
pathways (glycolysis) \cite{gho} and even individual (bio)chemical reactions 
(e.g. PO reaction) \cite{ols}.
In most of these cases, chaotic oscillations have been observed in 
addition to periodic behaviour. However, the route to chaos can 
be only studied in experimentally well tractable systems.

In addition to simple periodic and chaotic oscillations, complex 
periodic and quasiperiodic oscillations have been observed. 
Among the complex oscillations, bursting behaviour has been studied
thoroughly in the biological context, since it was observed in 
important processes such as nerve signal conduction \cite{wan} and signal 
transduction \cite{dix} within the cell. 
Several types of bursting have been characterized. 

Here, we study a recent model for calcium signal transduction \cite{Kummer}.
Calcium ions play a central role as second messengers within
the cell. Following the stimulation of agonist receptors at 
the cell membrane, a cascade of events is triggered which finally
leads to the massive liberation of calcium from intracellular stores. 
Calcium in turn influences multiple processes like gene expression, 
diverse enzymatic reactions, vesicular transport and embryogenesis
\cite{B1,B2}.

In agreement with experimental results, the model displays
simple periodic (spiking), bursting and chaotic oscillations.
We compare the bursting behaviour of the complete model to an independent
dynamics of its slow and fast subsystems.  
We study the bifurcation behaviour of the model with
one slow variable being used as the bifurcation parameter, considering
all other variables to be slaved.
This allows the classification of bursting, an approach first used
by J.\ Rinzel \cite{Rinzel} for bursting
electrical activity in models of nerve cells. 
We classify the bursting behaviour found in the model under investigation
to be of the Fold-Hopf type 
according to the comprehensive scheme proposed in \cite{Izhikevich}. 
So far, this type of bursting had
been described in the context of neural activity. Here, we report
the first observation in a model for calcium signal transduction. 
Reasons for this behaviour in terms of the connectivity of the system 
are discussed.

\begin{figure}
\begin{center}
 \epsfxsize=0.8\hsize \mbox{\hspace*{-.06 \hsize} \epsffile{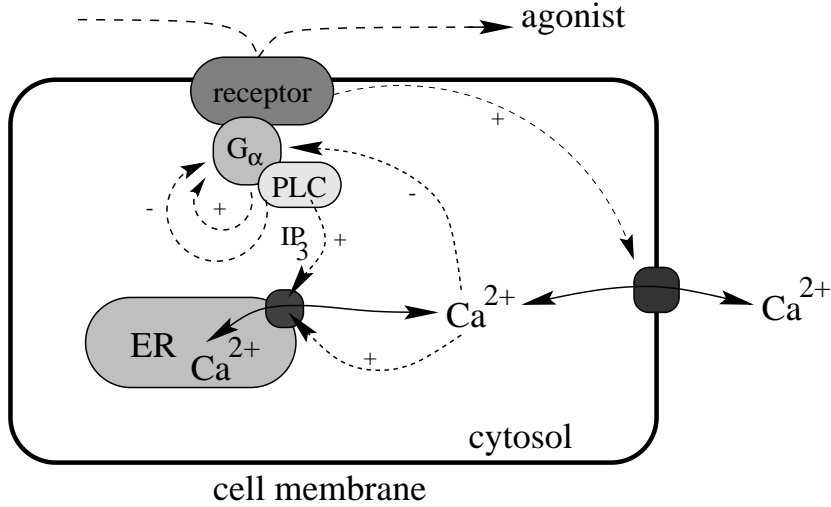} }
\end{center}
 \caption
 {Schematic drawing of the cell with a membrane bound receptor (top) and the
 endoplasmic reticulum (ER). 
 Dotted arrows denote the modelled influences on the Ca$^{2+}$ fluxes (solid arrows).
 }
 \label{cell}
\end{figure}

\section{Models for Calcium Dynamics}
\label{model}

Many calcium signal transduction models have been developed (for a review
see \cite{SKS95}). The one studied here and illustrated in Fig.~\ref{cell}
describes the following scenario:
Upon binding of the agonist, the G$_\alpha$-subunit 
of the receptor coupled G-protein is activated. Variable $a$ denotes the 
concentration of activated G$_\alpha$-subunit. 
This in turn activates phospholipase C. The concentration of the activated form 
PLC$^*$ is denoted by $b$. The activated G$_\alpha$-subunit 
and PLC$^*$ cause the liberation of calcium from the endoplasmic
reticulum (ER) and from the extracellular space. The calcium concentrations in 
the cytosol and in the endoplasmic reticulum are described by the variables
$c$ and $d$. Calcium is pumped back to the ER and to the extracellular space.
Moreover, different mechanisms lead to the inactivation of $a$ and $b$ 
over time. 

Variables $b$ and $c$ exhibit negative feedback on variable $a$ by 
acceleration of 
the inactivation processes. In addition, variables $a$ and
$c$ show autocatalytic behaviour \cite{Kummer}. The autocatalysis in $c$ is 
an integral compound of a term that describes the activity of a calcium 
channel at
the endoplasmic reticulum (ER). We modified this term compared to the original
model in order to make it more realistic. The change mainly ensures the
reversibility of calcium diffusion through the channel. The qualitative 
dynamics of the model is not affected by this change. 

The model thus has the following form:

\begin{eqnarray}
\frac{\mathrm{d}a}{\mathrm{d}t} &=& k_1+k_2 a-\frac{k_3 ab}{a+k_4}-\frac{k_5ac}{a+k_6}
\label{4.1}\\
\frac{\mathrm{d}b}{\mathrm{d}t} &=& k_7a-\frac{k_8b}{b+k_9}
\label{4.2}\\
\frac{\mathrm{d}c}{\mathrm{d}t} &=& \frac{k_{10}cb^4(d-c)}{b^4+k_{11}^4}+k_{12}b+k_{13}a-\frac{k_{14}c}{c+k_{15}}-\frac{k_{16}c}{c+k_{17}}
\label{4.3}\\
\frac{\mathrm{d}d}{\mathrm{d}t} &=& -\frac{k_{10}cb^4(d-c)}{b^4+k_{11}^4}+\frac{k_{16}c}{c+k_{17}}
\label{4.4}
\end{eqnarray}

This model exhibits spiking and bursting, as well as chaotic oscillations
in dependence on agonist stimulation that is proportional to the parameter $k_2$
\cite{Kummer}. 
Moreover, it can be reduced to a 3 variable model which still captures the 
essential dynamics of the 4 variable model. This reduced model has the 
following form:

\begin{eqnarray}
\frac{\mathrm{d}a}{\mathrm{d}t} &=& k_1+k_2a-\frac{k_3ab}{a+k_4}-\frac{k_5ac}{a+k_6}
\label{3.1}\\
\frac{\mathrm{d}b}{\mathrm{d}t} &=& k_7a-\frac{k_8b}{b+k_9}
\label{3.2}\\
\frac{\mathrm{d}c}{\mathrm{d}t} &=& k_{13}a-\frac{k_{14}c}{c+k_{15}}
\label{3.3}
\end{eqnarray}

Again, variables $b$ and $c$ exhibit negative feedback on $a$. 
The autocatalysis in 
$a$ is still present. However, the autocatalysis in $c$ is not necessary.

\section{Methods}
\label{methods}

In order to perform a bifurcation and linear stability analysis, the
numerical continuation software AUTO97 \cite{auto97} is employed.
AUTO97 computes fixed points and limit cycles of sets of ordinary differential
equations. It calculates branches of these solutions under parameter variation, 
detects bifurcation points and allows to switch to a new branch emerging at a
bifurcation. It also monitors eigenvalues respectively Floquet multipliers of 
the solutions to obtain their linear stability.
This method is used for both the analysis of the complete system as well as for
its separated slow and fast subsystems.

The classification of the bursting behaviour is carried out according to 
the following procedure: The bursting behaviour consists of two consecutive 
phases, a silent phase with slow changes of the variables and 
an active phase with rapid oscillations of some of the variables whereas 
others remain slowly varying. Hence, during the active phase, the 
individual variables change on different 
time scales and may be grouped into a slow and a fast subsystem, respectively
\cite{Rinzel}.
The encountered bifurcations of the fast subsystem that initiate and terminate
the active phase classify the type of bursting according to \cite{Izhikevich}.

\begin{figure}
\begin{center}
 \epsfxsize=0.95\hsize \mbox{\hspace*{-.06 \hsize} \epsffile{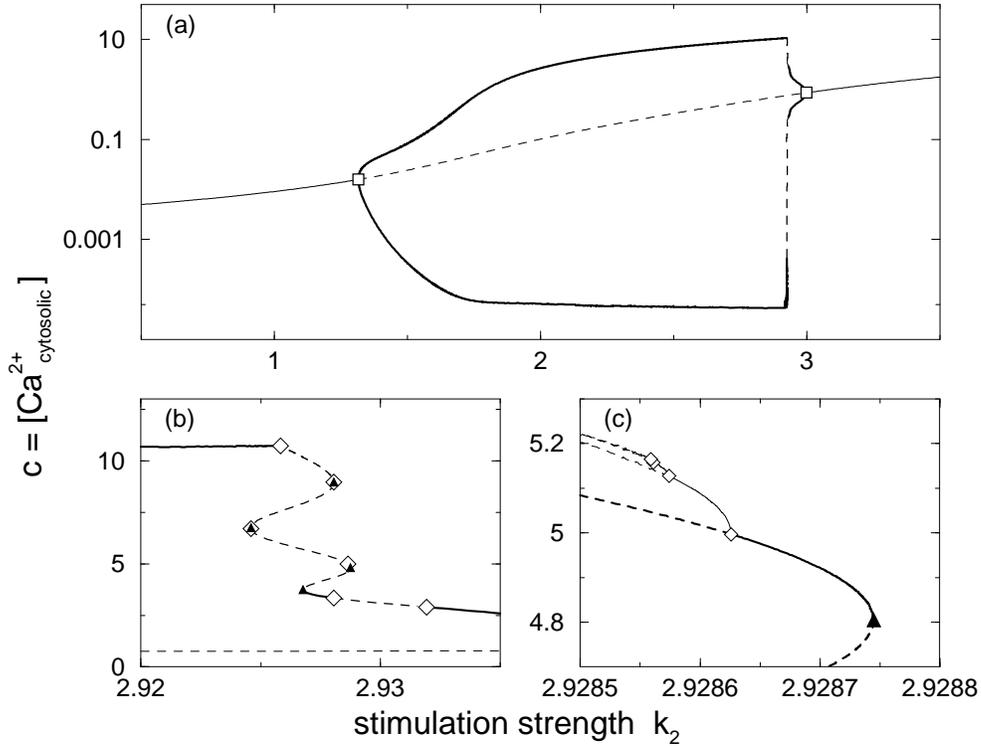} }
\end{center}
 \caption
 {Bifurcation diagrams of complex calcium oscillations (thick curves show maximum
 and minimum $c(t)$) and steady state behaviour (thin curves in (a,b)) 
 in Eqs.~(\ref{3.1})-(\ref{3.3}) as function of the agonist level $k_2$.
 Solid (dashed) curves give stable (unstable) solutions.
 Symbols denote Hopf (open square), saddle-node of periodic orbits 
 (filled triangle) and period doubling (open diamond) bifurcations.
 (a) covers the whole parameter range of interest whereas
 (b) is a close-up of the steady state (thin) and the maximum $c$ (thick
 curve) in the parameter range where periodic bursting is
 unstable and chaotic behaviour has been observed, e.g. at $k_2=2.9259$ in Fig.7 
 of \cite{Kummer}.
 (c) shows one of the period doubling cascades with 
 branches of 1 (thick), 2, 4 and 8 irregular bursts (thin) per period. 
 See the text for details.
 Parameters
 $k_1=0.212, 
 k_3=1.52, 
 k_4=0.19,
 k_5=4.88,
 k_6=1.18,
 k_7=1.24,
 k_8=32.24,
 k_9=29.09,
 k_{13}=13.58,
 k_{14}=153,
 k_{15}=0.16$ are the same as in Fig.7 of \cite{Kummer}.
 }
 \label{bif3}
\end{figure}

\section{Results}
\label{results}

First, we will focus on the 3 variable model (\ref{3.1})-(\ref{3.3})
and later compare its bifurcation structure to the 4 variable model.
We choose $k_2$ as the bifurcation parameter. It denotes the strength 
of agonist stimulation of the receptor and it is experimentally well studied.
All other parameters are kept fixed.
The bifurcation diagrams in Fig.~\ref{bif3} show the maximum and minimum of 
$c(t)$ versus the parameter $k_2$. 

\begin{figure}
\begin{center}
 \epsfxsize=0.95\hsize \mbox{\hspace*{-.06 \hsize} \epsffile{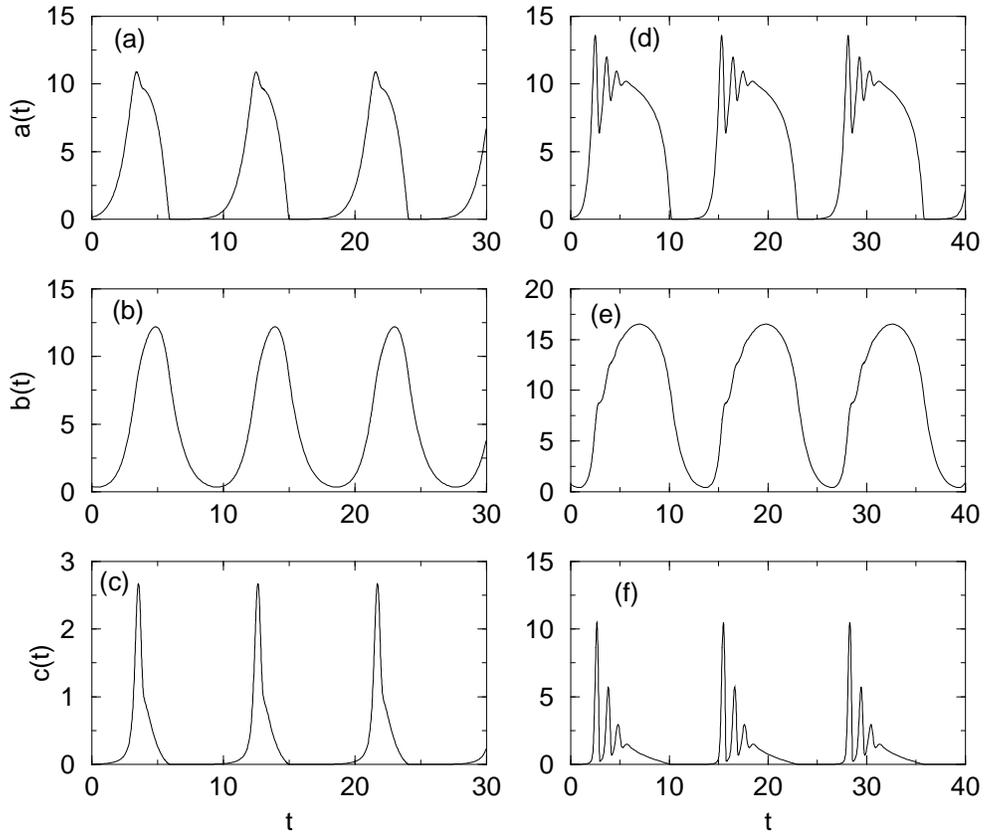} }
\end{center}
 \caption
 {Examples of spiking (a)-(c) at $k_2=2.0$ and bursting (d)-(f) at $k_2=2.9$
 in Eqs~(\ref{3.1})-(\ref{3.3}).
 The ordinate labels apply to both panels in each row.
 The bursts (d)-(f) consist of a silent phase with low values of $a(t)$ and
 $c(t)$ while $b(t)$ decreases and an active phase with rapid oscillations of 
 $a(t)$ and $c(t)$ when $b(t)$ monotonously increases.
 The dynamics of $a(t)$ and $c(t)$ occur on a fast and the dynamics of
 $b(t)$ on a slow time scale.
 Parameters are the same as in Fig.~\ref{bif3}.
 }
 \label{portraits}
\end{figure}

At $k_2<1.3$, the steady state is stable and $c$ increases monotonically 
with the stimulation $k_2$.
At $k_2=1.3$, a supercritical Hopf bifurcation renders the steady state unstable 
and stable periodic oscillations coexist with the unstable steady state for
$k_2>1.3$. 
As $k_2$ is increased in the interval $1.3<k_2<2.92$ the oscillations become
increasingly complex and develop secondary maxima in $c(t)$.
In Fig.~\ref{portraits} examples of spiking (a)-(c) at $k_2=2.0$ and bursting
(d)-(f) behaviour at $k_2=2.9$ are compared.

At $k_2=2.9258$, the branch of stable bursting undergoes a period doubling
bifurcation and turns unstable.
Stable oscillations are recovered after a sequence of saddle-node
and period doubling bifurcations of periodic orbits 
as shown in Fig.~\ref{bif3}(b).
Each hysteresis reduces the number of secondary spikes per burst by one.
The intervals of increasing maximum of $c(t)$ with $k_2$ are 
unstable due to the saddle-node bifurcations. Those of decreasing 
maximum are destabilized by pairs of period doubling cascades that occur very
close to the saddle-node bifurcations 
(compare panels (b) and (c) in Fig.~\ref{bif3}).

Fig.~\ref{bif3}(c) gives a close-up of a period doubling cascade from 
Fig.~\ref{bif3}(b) that is 
the reason for chaotic behaviour within the parameter intervals between pairs 
of such cascades. We only show 4 period doubling bifurcations with branches of 
1 (thick), 2, 4 and 8 bursts per period. Due to increasing computational effort,
the emerging branch with 16 irregular bursts per period is not computed further.
This chaotic behaviour has been reported earlier at $k_2=2.9259$ 
(Fig.7 in \cite{Kummer}). 

After the periodic oscillations changed back to simple spiking they terminate in
the second supercritical Hopf bifurcation at $k_2=3.0$.
For $k_2>3.0$, the steady state is
stable and corresponds to over-stimulation with constant high calcium 
concentration in the cytosol.
At large $k_2$, the model displays a third Hopf bifurcation where again
stable periodic oscillations emerge. However, this feature lies outside the
parameter range of interest.

\begin{figure}
\begin{center}
 \epsfxsize=0.95\hsize \mbox{\hspace*{-.06 \hsize} \epsffile{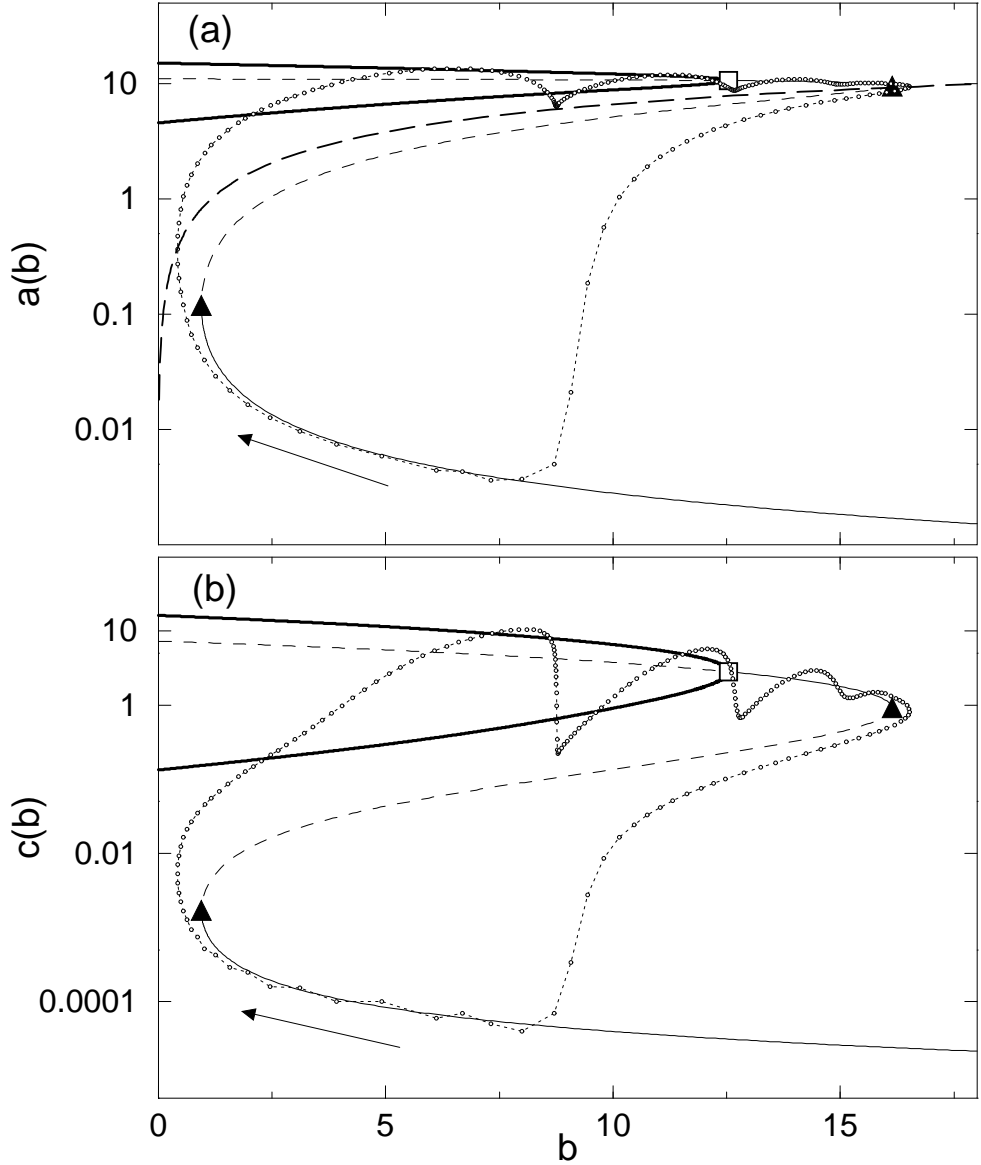} }
\end{center}
 \caption
 {Solutions of the fast subsystem $(a(t,b),c(t,b))$ at $k_2=2.9$ as in
 Fig.~\ref{portraits}(d)-(f), projected onto the $(b,a)$ and the $(b,c)$ 
 plane.
 Thin (thick) curves stand for steady states 
 (maxima and minima of fast oscillations) and dashed
 (full) curves indicate unstable (stable) solutions.
 The dotted curve (with circles) represents the projection of the solution of 
 the full system (\ref{3.1})-(\ref{3.3}) which is 
 parametrized by time in the clockwise direction following the arrow.
 Small irregularities at low $a$ and $b$ reflect the finite numerical accuracy.
 The thick dashed curve in (a) represents the nullcline $b_t=0$ of the slow 
 dynamics. $b(t)$ increases (decreases) above (below) this curve. 
 Parameters are the same as in Fig.~\ref{bif3}.
 }
 \label{slow}
\end{figure}

From Fig.~\ref{portraits}(d)-(f) it is evident that the smooth oscillations of 
$b(t)$ evolve on a slower time scale than the secondary spikes of $a(t)$ and 
$c(t)$. The same solution is given by the dotted curve (with circles) in 
Fig.~\ref{slow}.
Here, the dynamics in the three-dimensional phase space $(a,b,c)$ is displayed 
by projection onto the $(b,a)$ and $(b,c)$ planes.

In the following, the slow subsystem Eq.~(\ref{3.2})
and the fast subsystem Eqs.~(\ref{3.1}),(\ref{3.3}) are studied separately.
Setting Eq.~(\ref{3.2}) equal to zero one derives the nullcline 
\begin{equation}
a_0(b)=\frac{k_8/k_7 b}{b+k_9}
\label{a0}
\end{equation}
where d$b/$d$t\!>\!0$ for $a\!>\!a_0(b)$ and d$b/$d$t\!<\!0$ for
$a\!<\!a_0(b)$ as follows from the first term on the right hand side of 
Eq.~(\ref{3.2}).

The bifurcation diagrams of the fast subsystem Eqs.~(\ref{3.1}),(\ref{3.3}) 
with $b$ as bifurcation parameter are superimposed in Fig.~\ref{slow}.
As $b$ is varied, the steady state solution (thin curves) for $a$ and $c$ shows 
a hysteresis limited by two saddle-node bifurcations (filled triangles).
The lower branch is stable and the middle branch unstable. 
On the upper branch, a Hopf bifurcation (open square) gives raise to fast 
oscillations.
In the present model, the time scale separation is finite and the solution of
the full system (\ref{3.1})-(\ref{3.3}) slightly deviates from the curves 
derived from the fast subsystem.

We gain a qualitative understanding of the complex structure of
bursting behaviour if we follow the circles in clockwise direction (arrow) 
as time progresses.
Starting in the silent phase of the burst where $a$ and $c$ are low (compare
Fig.~\ref{portraits}(d)-(f)) the dynamics are close to the lower stable branch 
of $a(b)$ and $c(b)$. d$b/$d$t<0$ holds, since the nullcline 
(thick dashed curve) lies above.
Hence, $b$ will slowly decrease and $a$ and $c$ will adiabatically follow the
lower branch in Fig.~\ref{slow} until the lower saddle-node bifurcation 
is exceeded. Then, no stable steady state exists any more and 
the dynamics approach the fast oscillation (thick curve) at larger values of 
$a$ and $c$. Now $a>a_0$ and $b$ increases. 

The crossing of the saddle-node bifurcation triggers the active phase of 
the burst
which starts with a large spike in $a(t),c(t)$ followed by secondary spikes of 
decreasing amplitude. The spikes correspond to the branch of fast oscillations 
in Fig.~\ref{slow}.
At larger $b$ these oscillations become smaller and continuously vanish
in the supercritical Hopf bifurcation. 
However, since the time scale separation is finite, the complete dynamics passes
the Hopf bifurcation and returns via damped oscillations to the stable focus on
the upper branch. 
This feature has been analysed in detail in a model of enzyme kinetics 
that shows the same type of bursting \cite{Holden}.
After a short plateau (upper stable steady state) the saddle-node bifurcation 
at large $b$ is exceeded.
The lower steady state is approached and $b$ starts to decrease.
This completes one period of the burst.

Since the active phase starts at a saddle-node bifurcation (fold) and ends 
on a branch of stable foci provided by
a supercritical Hopf bifurcation, the present dynamics is an example of
Fold-Hopf bursting following the classification by 
E. M. Izhikevich \cite{Izhikevich}.

\begin{figure}
\begin{center}
 \epsfxsize=0.95\hsize \mbox{\hspace*{-.06 \hsize} \epsffile{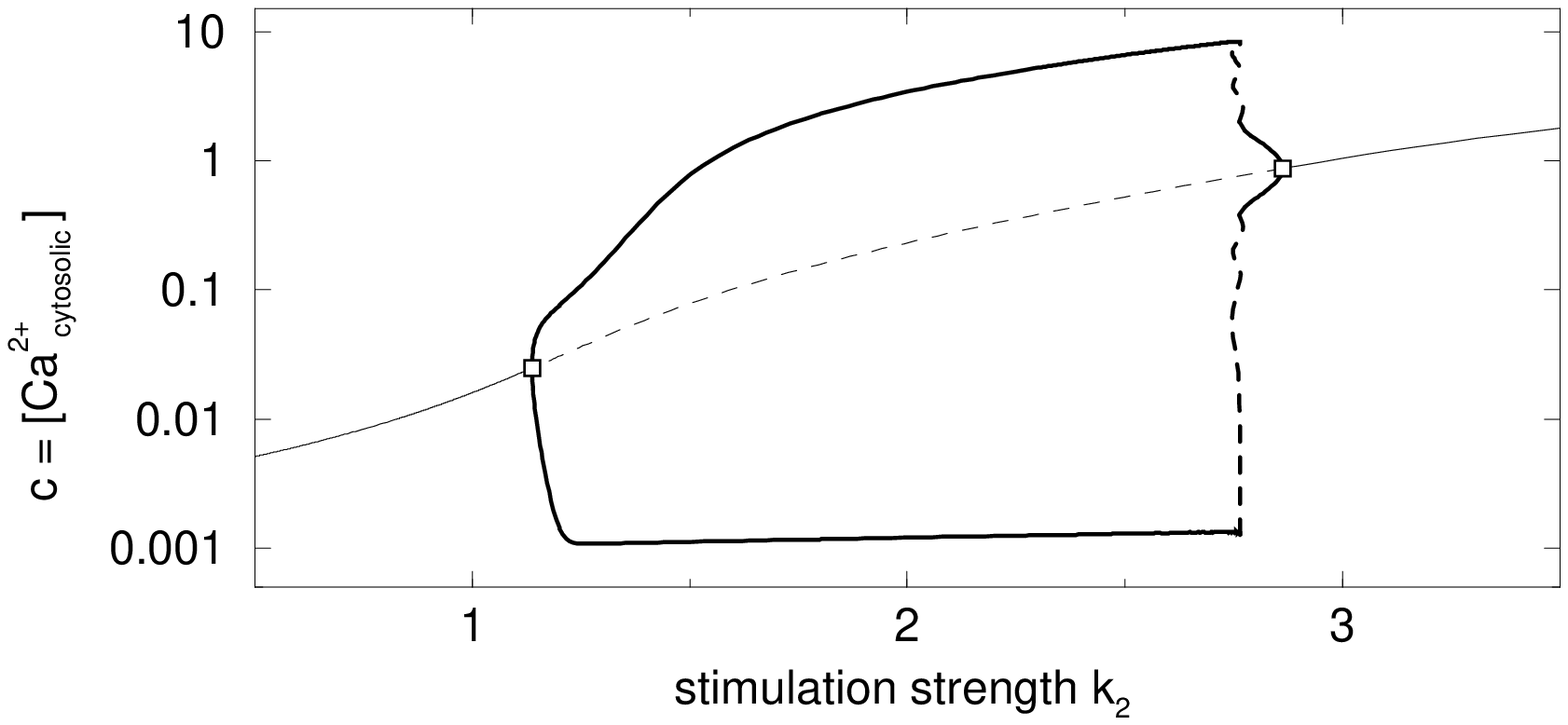} }
\end{center}
 \caption
 {The bifurcation diagram of the 4 variable model (\ref{4.1})-(\ref{4.4})
 is very similar to that of 
 the 3 variable model (\ref{3.1})-(\ref{3.3}).
 The branch of unstable oscillations (thick dashed) has the same structure as in
 Fig.~\ref{bif3}~.
 Curves have the same interpretation as in Fig.~\ref{bif3} and parameters are
 $k_1=0.09, 
 k_3=0.64, 
 k_4=0.19,
 k_5=4.88,
 k_6=1.18,
 k_7=2.08,
 k_8=32.24,
 k_9=29.09,
 k_{10}=5.0,
 k_{11}=2.67,
 k_{12}=0.7,
 k_{13}=13.58,
 k_{14}=153,
 k_{15}=0.16,
 k_{16}=4.85,
 k_{17}=0.05$~.
 }
 \label{bif4and5}
\end{figure}

We have studied the 4 variable model (\ref{4.1})-(\ref{4.4}) 
in the same way as the 3 variable model (\ref{3.1})-(\ref{3.3}).
The two models qualitatively show the same behaviour.
In Fig.~\ref{bif4and5} we show the bifurcation diagram of the 4 variable model. 
The solutions have a similar shape
and the bifurcation diagrams of the fast subsystems are qualitatively the same.
In the two models $b(t)$ is the only slow variable and the remaining
variables constitute the fast subsystem.
The type of bursting is the same in the two models.

\section{Discussion}
\label{discus}

We studied the bifurcation and bursting behaviour of a recently proposed model
for calcium signal transduction and classified the bursting behaviour
to be of Fold-Hopf type. This Fold-Hopf type of bursting has also been 
observed in the electrical activity of pyramidal cells of the cat hippocampus 
\cite{Kandel} and in models of the bursting electrical activity in 
pancreatic $\beta$-cells \cite{Smolen}. The authors of the latter paper 
call it ``tapered'' bursting. The models have been reduced to the Li\'enard 
form which has been studied intensively \cite{deVries,Pernarowski}. 
The same type of bursting was reported and called ``Type V'' \cite{deVries}.
However, the classification scheme for bursting by E. M. Izhikevich 
\cite{Izhikevich}
follows a bottom-up approach and provides a systematic nomenclature.
It is therefore preferred by us.

In the purely biochemical context, the only previous example for Fold-Hopf 
type bursting is an abstract enzymatic system \cite{dec}. 
This system is also subject of the study mentioned above \cite{Holden}. 
It can be described by three variables
and has the following form:

\begin{eqnarray}
\alpha' &=& \kappa - \delta\phi(\alpha, \beta) \label{enz1} \\
\beta'&=& q_1\delta\phi(\alpha, \beta) - \delta\nu(\beta,\gamma) \label{enz2} \\
\gamma' &=& q_2\delta\nu(\beta, \gamma) - k_s\gamma \label{enz3}
\end{eqnarray} 

It is interesting to compare the topology of this abstract biochemical 
system with the topology of the calcium signal transduction system studied
here. The abstract system involves two negative feedback loops of 
variable $\beta$ on $\alpha$ as well as $\gamma$ on $\beta$. 
It also contains two autocatalysis of the variables $\gamma$ and 
$\beta$. 

The 4 variable model studied here also contains two negative feedback loops as
well as two autocatalytic species. However, the 3 variable model which displays
the same dynamics contains only one autocatalytic species. 
Therefore, the overall topology of the two systems is similar but not the same. 

In \cite{kaern}, the model (\ref{enz1})-(\ref{enz3}) has been extended by two 
quickly relaxing variables that account for the finite time scale of allosteric 
transitions in the enzymatic system. 
The bursting behaviour was shown to depend on the chosen separation of time 
scales with the original dynamics (\ref{enz1})-(\ref{enz3}) being conserved in 
the limit of fast allosteric transitions.
Our 3 variable and 4 variable models also assume fast allosteric transitions
and so far other experimental indications are not available.
On the other hand, our models possess a different topology and the results 
for slow allosteric transitions \cite{kaern} do not immediately apply to the 
present situation.

In general, negative feedback loops and simple autocatalysis are very common
in biochemical systems. Cooperative or allosteric  
behaviour in enzymatic systems can be considered to be an autocatalysis 
for certain ranges of parameters. This behaviour is involved in 
metabolic regulation 
and signal transduction. It offers the possibility to fine tune 
biochemical systems, but also to display a wealth of different 
dynamic properties. Since the basic requirements for this behaviour
are so abundant in the biochemical network in living cells, 
it is likely that there are many more examples which have not been
discovered yet. Bursting behaviour in biology seems to be an integral 
part of the information processing machinery.   

How information is processed in the cell is still not very well understood.
However, it has been studied in some detail in the context of 
calcium signal transduction. It has been shown that the frequency of 
calcium oscillations encodes information \cite{li,dol,dek,oan}. 
Fundamental differences
in the topology of the signal transduction pathway also offer the 
possibility to give rise to fundamentally different dynamical behaviour and
encode information in this way \cite{Kummer}.  
In this context, it is of importance which type of bursting behaviour is 
displayed. Upon variation of the model parameters the locations of bifurcation 
points shift along the steady state branches in the fast subsystem.
For different types of bursting these quantitative changes may trigger different
qualitative changes of the bursting behaviour.

Other models of calcium dynamics have recently been studied that also show
bursting behaviour but of Fold-Sub-Hopf hysteresis \cite{Gold01}
and Sub-Hopf-Fold-Cycle type \cite{Habe}, respectively.
In these models the amplitude of the secondary spikes increases as the end of 
the active phase is approached. 
This corresponds to unstable foci in their fast
subsystems that are provided by subcritical Hopf bifurcations.
With our model, the secondary spikes decrease in amplitude.
More detailed experimental studies in the future will reveal which type(s) 
of bursting are predominant in calcium signal transduction.\\

\noindent{\bf Acknowledgements:}

UK and MO thank the Klaus Tschira Foundation for funding.


\begin{thebibliography}{999}

\bibitem{beg} M.\ J.\ Begon, L.\ Harper and C.\ R.\ Townsend, Ecology:
Individuals, Populations, and Communities, 3rd edition, Blackwell Science Ltd.\
Cambridge, MA (1996).

\bibitem{edm} L.\ N.\ Edmunds (ed.) Cell Cycle Clocks, Marcel Dekker, 
New York (1984).

\bibitem{gho} A.\ Ghosh and B.\ Chance, Biochem.\ Biophys.\ Res.\ Commun.\ {\bf 16}
(1964) 174.

\bibitem{ols} L.\ F.\ Olsen and H.\ Degn, Nature {\bf 267} (1977) 177.

\bibitem{wan} X.\ J.\ Wang and J.\ Rinzel, Brain Theory and Neural Networks
(M.\ A.\ Arbib, ed.), The MIT press, Cambridge, MA (1995).

\bibitem{dix} C.\ J.\ Dixon, N.\ M.\ Woods, T.\ E.\ Webb and A.\ K.\ Green, 
Biochem.\ J.\ {\bf 269} (1990) 499.

\bibitem{Kummer} U.\ Kummer, L.\ F.\ Olsen, C.\ J.\ Dixon, A.\ K.\ Green, E.\
Bornberg-Bauer and G.\ Baier, Biophys.\ J.\ {\bf 79} (2000) 1188. 

\bibitem{B1} M.\ J.\ Berridge, Nature {\bf 361} (1993) 315.

\bibitem{B2} M.\ J.\ Berridge, M.\ D.\ Bootman and P.\ Lipp, Nature {\bf 395} 
(1998) 645.

\bibitem{Rinzel} J.\ Rinzel, in Mathematical topics in population biology,
morphogenesis, and neurosciences, Eds.\ E.\ Teramoto and M.\ Yamaguti,
Springer, Berlin, (1987).

\bibitem{Izhikevich} E.\ M.\ Izhikevich, Int.\ J.\ Bifurcat.\ Chaos {\bf 10} 
(2000) 1171.

\bibitem{SKS95} J.\ Sneyd, J.\ Keizer and M.\ J.\ Sanderson, FASEB J.\ {\bf 9} 
(1995) 1463. 

\bibitem{auto97} E.\ Doedel, A.\ Champneys, T.\ Fairgrieve, Y.\ Kusntsov, 
{\it et al.}, {\sc AUTO97}:{\it Continuation and bifurcation software for ordinary
differential equations} (Concordia University, Montreal, 1997).

\bibitem{Holden} L.\ Holden and T.\ Erneux, 
SIAM J.\ Appl.\ Math.\ {\bf 53} (1993) 1045.

\bibitem{Kandel} E.\ R.\ Kandel and W.\ A.\ Spencer,
J.\ Neurophysiol.\ {\bf 24} (1961) 243.

\bibitem{Smolen} P.\ Smolen, D.\ Terman and J.\ Rinzel,
SIAM J.\ Appl.\ Math.\ {\bf 53} (1993) 861.

\bibitem{deVries} G.\ de Vries, J.\ Nonlinear Sci.\ {\bf 8} (1998) 281.

\bibitem{Pernarowski} M.\ Pernarowski,
SIAM J.\ Appl.\ Math.\ {\bf 54} (1994) 814.

\bibitem{dec} O.\ Decroly and A.\ Goldbeter, J.\ Theoret.\ Biol.\ {\bf 124} (1987) 219.

\bibitem{kaern} M.\ K{\ae}rn and A.\ Hunding, J.\ Theoret.\ Biol.\ {\bf 198} (1999) 269.

\bibitem{li} W.\ Li, J.\ Llopis, M.\ Whitney, G.\ Zlokarnik and R.\ Y.\ Tsien, Nature {\bf 392} (1998) 936.

\bibitem{dol} R.\ E.\ Dolmetsch, K.\ Xu and R.\ S.\ Lewis, Nature {\bf 392} (1998) 933.

\bibitem{dek} P.\ De Koninck and H.\ Schulman, Science {\bf 279} (1998) 227.

\bibitem{oan} E.\ Oancea and T.\ Meyer, Cell {\bf 95} (1998) 307.

\bibitem{Gold01} J.\ A.\ M.\ Borghans, G.\ Dupont and A.\ Goldbeter, 
Biophys.\ Chem.\ {\bf 66} (1997) 25.

\bibitem{Habe} T.\ Haberichter, M.\ Marhl and R.\ Heinrich,
Biophys.\ Chem.\ {\bf 90} (2001) 17.

\end{thebibliography}
\end{document}